\documentclass[aps,prf,reprint,notitlepage,onecolumn,superscriptaddress]{revtex4-1}

\usepackage{amsmath,amssymb}
\usepackage{graphicx}
\usepackage{subcaption}
\usepackage{bm}
\usepackage{epstopdf}
\usepackage{verbatim} 
\usepackage[normalem]{ulem}

\usepackage{xcolor}
\definecolor{light-gray}{gray}{0.5}
\definecolor{blue}{rgb}{0.0,0.0,1.0}
\definecolor{green}{rgb}{0.0,0.5,0.0}
\definecolor{red}{rgb}{1.0,0.0,0.0}
\definecolor{cyan}{rgb}{0.0,0.75,0.75}
\definecolor{magenta}{rgb}{0.75,0.0,0.75}
\definecolor{yellow}{rgb}{0.75,0.75,0.0}
\definecolor{orange}{rgb}{0.9,0.3,0.0}

\newcommand{\grad}{\bm \nabla}
\newcommand{\pd}{\partial}

\newcommand{\sk}[1]{\textcolor{black}{#1}}

\newcommand{\lhat}{\widehat}

\begin{document}

\title{Bifurcations of a 
plane parallel flow with Kolmogorov forcing}

\author{Kannabiran Seshasayanan}
\email[]{s.kannabiran@gmail.com}
\affiliation{Universit{\'e} Paris-Saclay, CEA, CNRS, SPEC, 91191, Gif-sur-Yvette, France}
\author{Vassilios Dallas}
\email[]{vassilios.dallas@gmail.com}
\affiliation{Mathematical Institute, University of Oxford, Woodstock Road, Oxford OX2 6GG, UK}
\author{Stephan Fauve}
\email[]{fauve@lps.ens.fr}
\affiliation{Laboratoire de Physique de l'\'Ecole normale sup\'erieure, ENS, Universit\'e PSL, CNRS, Sorbonne Universit\'e, Universit\'e Paris-Diderot, Sorbonne Paris Cit\'e, Paris}

\begin{abstract}
We study the primary bifurcations of a two-dimensional Kolmogorov flow in a channel subject to boundary conditions chosen to mimic a parallel flow, i.e. periodic and free-slip boundary conditions in the streamwise and spanwise directions, respectively. The control parameter is the Reynolds number based on the friction coefficient, denoted as $Rh$. We find that as we increase $Rh$ the laminar steady flow goes through a degenerate Hopf bifurcation with both the oscillation frequency and the amplitude of the \sk{growing} mode being zero at the threshold. A reduced four-mode model captures the scalings that are obtained from the numerical simulations. As we increase $Rh$ further we observe a secondary instability which excites the largest mode in the domain. The saturated amplitude of the largest mode is found to scale as a $3/2$ power-law of the distance to the threshold which is also explained using a low-dimensional model. 
\end{abstract}

\maketitle


\section{Introduction} 
\label{sec:intro}

The two-dimensional flow in a doubly periodic domain driven by a Sine wave body forcing was first introduced in 1959 by Kolmogorov \cite{arnoldmeshalkin60} as a mathematically tractable problem to study flow stability. It has been shown that this flow is unstable above a critical Reynolds number of order one in the limit of an unbounded flow domain \cite{meshalkin1961}. The instability occurs at vanishing wave number which has been used to perform a weakly nonlinear analysis showing that a large scale flow is generated through a stationary pitchfork bifurcation \cite{nepomniashchii1976,sivashinsky1985weak,lucaskerswell14}.

Experiments on Kolmogorov flows were first carried out using thin layers of electrolytes \cite{bondarenko1979laboratory} or liquid metals \cite{sommeria86} with spatially periodic driving by the Lorentz force and more recently in soap films with hydrodynamic driving \cite{burgessetal99}. It was realised that in all realistic configurations, a linear friction force should be added to the two-dimensional Navier-Stokes equation in order to model the experimental results. In the case of a one-dimensional spatial forcing, this friction term inhibits the large scale flow such that the first instability occurs at finite wave number.

Another important aspect concerns the effect of boundary conditions. Mixed boundary conditions have been used in order to mimic experimental configurations. Periodic boundary conditions have been kept in the streamwise direction whereas stress-free boundary conditions have been used in the spanwise direction. This lateral confinement of the base flow suppresses the instability at vanishing wave number even in the absence of linear friction \cite{thess92,fukutamurakami98}. The instability comes in at finite wave number although the wave number decreases when the confinement length $L$ is increased \cite{fukutamurakami98}. The first instability threshold decreases to the value 
of the unbounded flow in the limit of $L \gg 1$.  More surprisingly, the nature of the 
primary bifurcation depends on the confinement. It has been first experimentally observed that in the case of strong confinement, when only half wave length of the base flow fits in the channel ($N=1$), the first instability is oscillatory \cite{kolesnikov1985} whereas it is stationary for $N=6$ \cite{bondarenko1979laboratory}. It has been observed later that the nature of the bifurcation depends on the parity of $N$ \cite{batchaev1989}. Traveling waves are generated when $N$ is odd, whereas a stationary regime is observed when $N$ is even except for $N=4$ for which an oscillatory regime is found. Linear stability analysis confirmed that the value of $N$ affects the nature of the bifurcation. A Hopf bifurcation occurs for $N=2$ whereas it is stationary for $N=4$ and $N=6$ \cite{thess92}. This is not in agreement with the experiments but we note that 
the lateral boundary conditions are different.
The nature of the bifurcation with respect to the flow confinement has been carefully analysed \cite{chen2002,chenprice05,chen20} but no simple argument has been put forward. 
Note that the definition of $N$ in \cite{thess92} is based on the number of the wavelengths instead of the number of half wavelengths as defined here.

In 
this study we show that even though the growth rate of the first instability is real for $N=4$, the bifurcation is not, strictly speaking, a stationary one but is a degenerate Hopf bifurcation. Indeed, a limit cycle is generated but its frequency vanishes at the instability onset. This process does not belong to one of the generic bifurcation scenarios that generate a limit cycle.  A supercritical Hopf bifurcation occurs at vanishing amplitude but finite frequency. In contrast, a limit cycle can be generated with finite amplitude and infinite period when two fixed points on an invariant cycle undergo a saddle-node bifurcation and disappear or when a limit cycle collides with a saddle point leading to a homoclinic bifurcation \cite{guckenheimer1983}. In our case, both the amplitude and the frequency of the limit cycle vanish at threshold. We understand this behaviour using a reduced set of interacting triads in section  \ref{sec:12bif_the}.

As recalled above, in the case of a one-dimensional forcing, fluid friction as well as lateral confinement of the flow prevent the generation of a large scale flow at the primary instability threshold.  However, in two-dimensional forcing configurations, it has been observed that a large scale shear flow can be generated by the first instability of a linear array of confined counter-rotating vortices \cite{tabeling1987}. This has been confirmed by numerical simulations \cite{guzdar1994} but a weakly nonlinear analysis of the type \cite{nepomniashchii1976,sivashinsky1985weak} is not possible in that case due to the boundary conditions. Above the primary 
instability mentioned before, the flow becomes two-dimensional and we could expect that a secondary bifurcation generates a large scale flow. This indeed occurs and a streamwise-independent shear flow with half of a wave length fitting in the channel is generated. Its amplitude increases above threshold with a $3/2$ power-law scaling which is at odd with respect to the characteristic behaviour of supercritical bifurcations. A $1/2$ power-law scaling is observed most of the time except in the vicinity of tricritical points for which the coefficient of cubic nonlinearities vanishes giving rise to a $1/4$ power-law scaling \cite{petrelis2005}. The $3/2$ power-law scaling results from the nonlinear forcing of the shear flow by modes that bifurcate at the secondary instability threshold. The large scale shear flow breaks mirror symmetry with respect to the mid-plane of the channel such that two mean flow solutions with opposite signs exist. When a turbulent regime is reached, random mean flow reversals are observed, which 
were recently studied in \cite{dallas2019abrupt}.

The article is organised as follows. In section \ref{sec:setup} we describe the flow configuration. 
In section \ref{sec:12bif_dns}, we present results about the first and the second bifurcations undergone by the system from direct numerical simulations (DNS) of the fully nonlinear system and from the eigenvalue problem of the linearised system.
Next in section \ref{sec:12bif_the} we explain the results obtained from DNS with the help of reduced models of interacting modes. Conclusions are presented in section \ref{sec:conclus}.

\section{Problem set-up\label{sec:setup}}
\begin{figure}[!ht]
  \vspace{0.3cm}
   \includegraphics[width=0.6\textwidth]{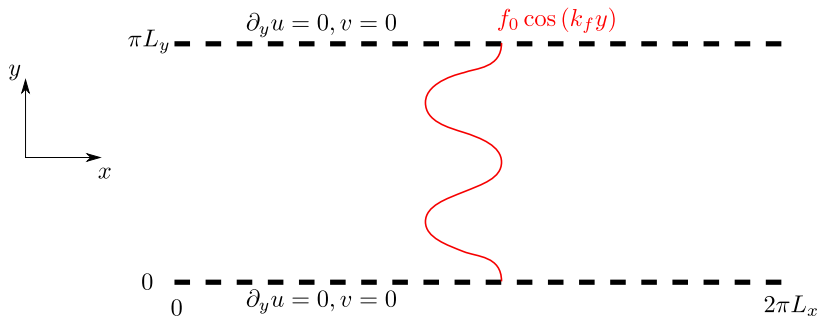}
 \caption{(Color online) Sketch of the domain under study. The red line represents the spatial form of the Kolmogorov forcing. 
 Note that $f_0\cos(k_f y)$ profile corresponds to the force that acts on $u$, the $x$-component of the velocity field.}
 \label{fig:domain}
 \end{figure}

We consider the two-dimensional Navier-Stokes equations for an incompressible velocity field ${\bf u} =  \grad \times \psi{\bf \hat z}$ forced by a Kolmogorov type forcing in a domain of extent 
$(x,y) \in \left[0, 2 \pi L_x \right] \times \left[0, \pi L_y \right]$ as illustrated in Fig. \ref{fig:domain}. The governing equation written in terms of the streamfunction $\psi (x,y,t)$ is given by
\begin{equation}
 \pd_t \psi + \grad^{-2}\{\grad^2\psi,\psi\} = \nu\grad^2\psi - \mu\psi + f_0\sin(k_f y),
 \label{eq:NS}
\end{equation}
where $\{f,g\} = f_x g_y - g_x f_y$ is the standard Poisson bracket (subscripts here denote differentiation), $\nu$ is the kinematic viscosity, $\mu$ is the friction coefficient, $f_0$ is the amplitude of the Kolmogorov forcing and $k_f$ is the forcing wave number. 
The boundary conditions are taken to be periodic in the $x$ direction and free-slip in the $y$ direction, i.e. $\psi = \psi_{yy} = 0$ at $y = 0, \pi L_y$.
%
We define the Reynolds number 
as $Re = f_0^{1/2} L_x /\nu$ and the friction Reynolds number as 
\begin{equation}
Rh = f_0^{1/2}/(\mu L_x).
\end{equation}
The control parameter of the problem is $Rh$ and we fix the Reynolds number to $Re = 1000$, the forcing wave number with respect to the height to $k_f L_y = 4$ and the aspect ratio of the domain to $2 \pi L_x/(\pi L_y) = 2$. For the rest of the article, all quantities are non-dimensionalised with the velocity scale $f_0^{1/2}$, the length scale $L_x$ and the time scale $L_x/f_0^{1/2}$. The scaling behaviour of the bifurcations we present in this article can be reproduced if one chooses the rms velocity as the relevant velocity scale instead of $f_0^{1/2}$. Our choice to non-dimensionalise using $f_0^{1/2}$ makes the analytical calculations more convenient.

We perform direct numerical simulations (DNS) by integrating Eq. \eqref{eq:NS} using the pseudospectral method \cite{gottlieborszag77,gomezetal05}. We decompose the streamfunction into basis functions with Fourier modes in the $x$ direction and Sine modes in the $y$ direction that satisfy the boundary conditions
\begin{equation}
 \psi(x,y,t) = \sum_{k_x=-\frac{N_x}{2}}^{\frac{N_x}{2}-1} \sum_{k_y=1}^{N_y} \widehat{\psi}_{_{k_x,k_y}}(t) \; e^{i k_x x} \sin(k_y y), \label{eqn:normalmodes}
\end{equation}
with $\widehat{\psi}_{_{k_x,k_y}}$ being the amplitude of the mode $\left(k_x, k_y \right)$ and $(N_x, N_y)$ denote the number of 
spectral modes in the $x, y$ coordinates respectively. 
For the streamfunction $\psi(x,y,t)$ to be real the following relation is satisfied in spectral space
\begin{equation}
  \label{eq:normal}
  \widehat{\psi}_{k_x, k_y} = \widehat{\psi}_{- k_x, k_y}^{*} .
\end{equation}
A third-order Runge-Kutta scheme is used for time advancement and the aliasing errors are removed with the two-thirds dealiasing rule which implies that the maximum wavenumbers are $k_x^{max} = N_x/3$ and $k_y^{max} = 2N_y/3$. 
The resolution was fixed to $(N_x,N_y) = (512,128)$ for all the simulations done in this study. The only simulations that required $512^2$ resolution were those with $k_f \geq 63$ (see Table \ref{tbl:kfLy}).


%
\section{ Primary and secondary bifurcations} \label{sec:12bif_dns}

\begin{figure}[!ht]
 \begin{subfigure}{0.49\textwidth}
   \includegraphics[width=\textwidth]{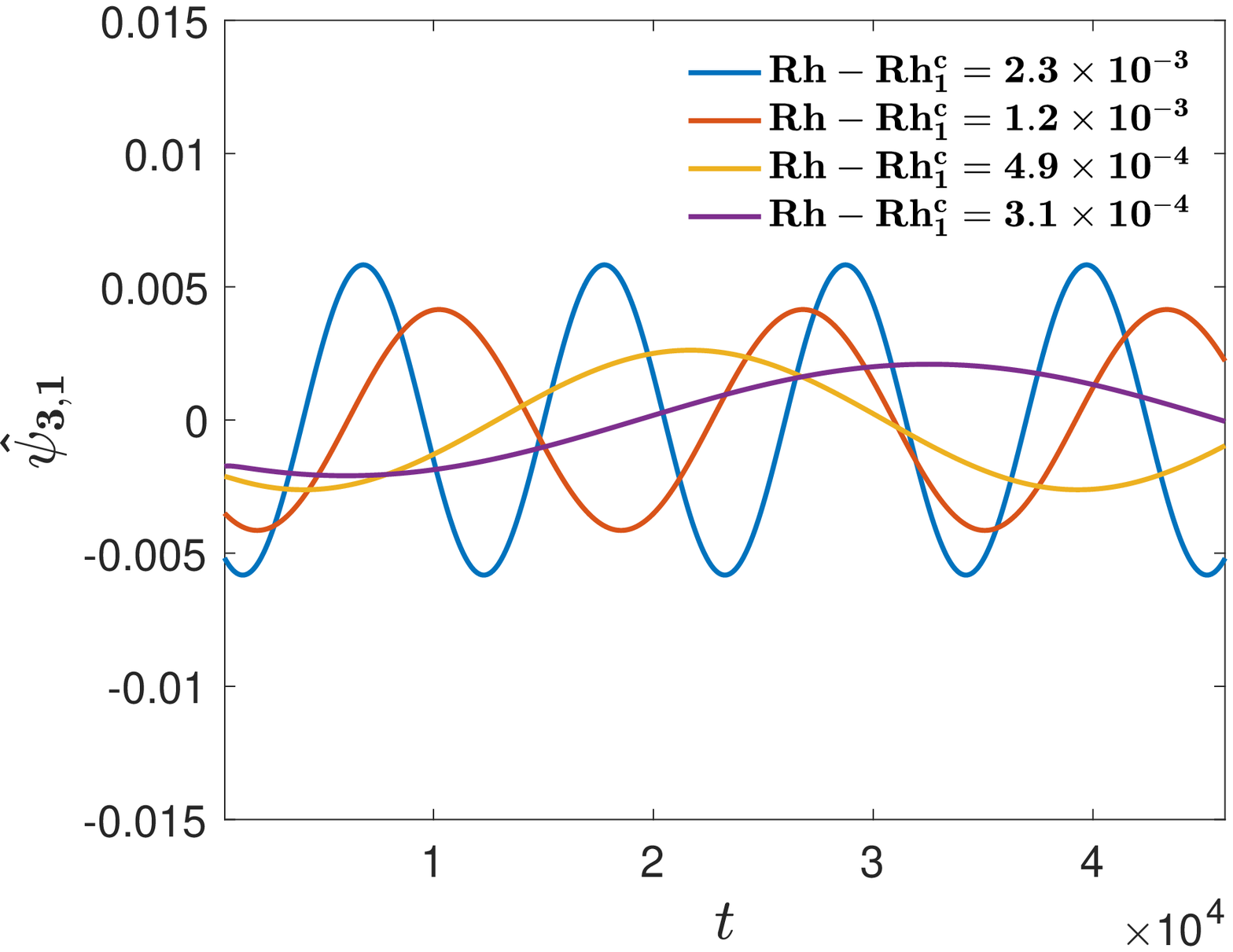}
   \caption{}
\label{fig:dns_time_series}
 \end{subfigure} 
 \begin{subfigure}{0.49\textwidth}
   \includegraphics[width=\textwidth]{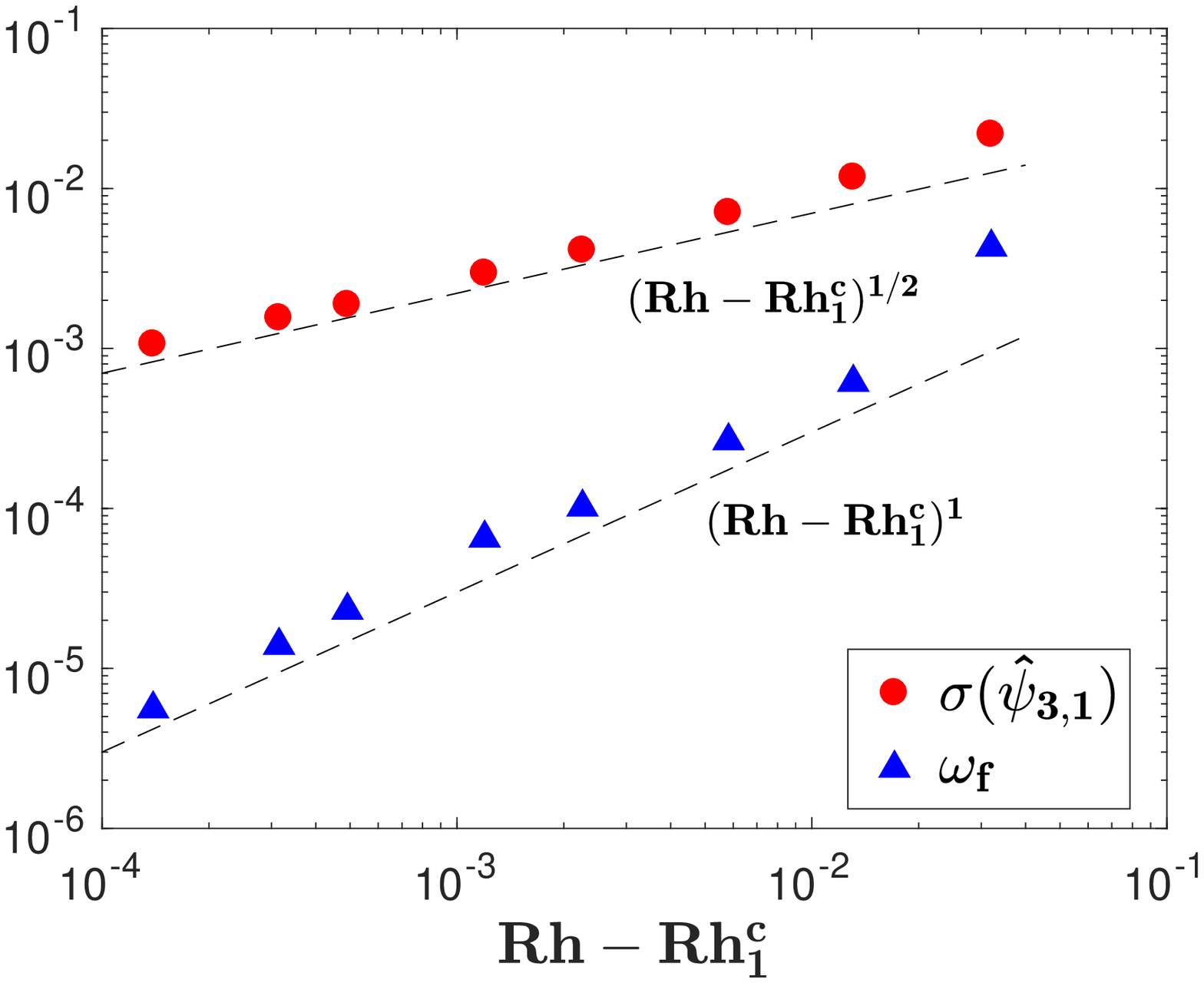}
   \caption{}
\label{fig:dns_scaling_first_hopf}
 \end{subfigure}
  \caption{(Color online) (a) Time series of the \sk{growing} mode $\widehat{\psi}_{3,1}$ for different values of $Rh$ close to the threshold. 
  (b) The standard deviation $\sigma ( \widehat{\psi}_{3,1} )$ and the oscillation frequency $\omega_f$ of the mode $\widehat{\psi}_{3,1}$ as a function of the distance to the threshold.}
 \label{fig:dns} 
 \end{figure}

For small values of $Rh$ a laminar flow is established which results from the balance between the forcing and the dissipation. Its expression is given by
\begin{align}
\psi \left(x, y, t \right) = \frac{1}{16 Re^{-1} +Rh^{-1}} \sin \left( 4 y \right). \label{eqn:baseflow}
\end{align}
This base flow corresponds to a parallel flow with the same spanwise structure as the forcing. We can represent the base flow in the Fourier-Sine basis Eq. \eqref{eqn:normalmodes}, which gives the only non-zero mode to be $\widehat{\psi}_{0,4} = 1/(16 Re^{-1} + Rh^{-1})$.  
From the DNS we observe that this laminar flow becomes linearly unstable above the critical value of
$Rh > Rh^c_1 \approx 0.593$, 
with the instability breaking the translational invariance in the $x$ direction. 
In Fig. \ref{fig:dns_time_series} we show the time series from the DNS of the most dominant Fourier-Sine mode $\widehat{\psi}_{3,1} (t)$ related to the instability, for different values of $Rh$ above the threshold. 
The time series demonstrate that as we approach the threshold
 $Rh - Rh_1^c \ll 1$,
both the amplitude and the oscillation frequency decrease. 
In Fig. \ref{fig:dns_scaling_first_hopf} we show the standard deviation $\sigma(\widehat{\psi}_{3,1})$ and the oscillation frequency $\omega_f$ of the saturated mode $\widehat{\psi}_{3,1}$ as a function of the distance to the threshold $Rh - Rh_1^c$. 
To capture the exact threshold we linearise around the base flow solution Eq. \eqref{eqn:baseflow} 
and we numerically solve the eigenvalue problem that arises from the linearised system of equations. 
The eigenvalue solver confirms the value of the threshold $Rh^c_1 \approx 0.593$ found from the DNS. The growing eigenmode has a wavenumber $k_x = 3$ in the $x$ direction and its projection on to Sine-basis in the $y$ direction shows that only the odd modes 
$k_y = 2 n + 1$, $n \in \mathbb{N}$ are excited.
In addition, it shows that the eigenvalues are real therefore leading to either an exponentially growing or decaying solution without any oscillatory behaviour. 
This is in agreement with the linear stability analysis of Thess \cite{thess92}. Therefore, the oscillations we observe result from the fully nonlinear problem, which is also responsible for the scaling of the frequency $\omega_f \propto Rh - Rh_1^c$. 
A four-mode model is presented in section \ref{sec:psi31model} to explain the observed behaviour. 

We study the nature of the first bifurcation by varying $k_f$ systematically, looking at both the linearised system and the fully nonlinear system. We remind that $k_f$ is non-dimensional and changing $k_f$ is equivalent to changing $k_f \, L_y$ in dimensional units.
In Table \ref{tbl:kfLy} we report the results we get from the eigenvalue problem for the linearised system and from the DNS.
\begin{table}[!ht]
\begin{tabular}{ |c|c|c|c|c| } 
\hline
$k_f$ & Linear problem & Nonlinear problem & $Rh_1^c$ & Largest amplitude \\ 
&  &  &  & \sk{growing} mode $(k_x,k_y)$ \\
\hline
\hline
2 & Hopf & Hopf & 1.428 & $(1,1)$ \\ 
\hline
3 & Hopf & Hopf & 1.077 & $(2,2)$ \\ 
\hline
4 & Pitchfork & Degenerate Hopf & 0.593 & $(3,1)$ \\ 
\hline
5 & Pitchfork & Pitchfork & 0.446 & $(3,1)$ \\ 
\hline
6 & Pitchfork &  Degenerate Hopf & 0.352 & $(4,1)$ \\ 
\hline
7 & Pitchfork &  Pitchfork & 0.299 & $(4,1)$ \\ 
\hline
8 & Pitchfork & Degenerate Hopf & 0.256 & {$(5,1)$} \\ 
\hline
9 & Pitchfork & Pitchfork  & 0.227 & {$(6,1)$} \\ 
\hline
10 & Pitchfork & Degenerate Hopf & 0.202 & {$(6,1)$} \\ 
\hline
63 & Pitchfork & Pitchfork & 0.0341 & $(36,1)$ \\ 
\hline
64 & Pitchfork & Degenerate Hopf & 0.0336 & $(37,1)$ \\ 
\hline
127 & Pitchfork & Pitchfork & 0.0191 & $(72,1)$ \\ 
\hline
128 & Pitchfork & Degenerate Hopf & 0.0190 & $(73,1)$ \\ 
\hline
\end{tabular}
\caption{The dependence of the nature of the first bifurcation on the forcing wavenumber $k_f$. The Reynolds number is fixed at $Re = 1000$ for all cases. The largest amplitude \sk{growing} mode shown in the last column is found from the eigenvalue problem. }
\label{tbl:kfLy}
\end{table}
 For $k_f = 2, 3$ both the linear and the non-linear problem give rise to a Hopf bifurcation. Then for $k_f \geq 4$ the linear problem gives rise to a pitchfork bifurcation, while the nonlinear problem gives a pitchfork only when $k_f$ is odd and a degenerate Hopf when $k_f$ is even. These results with odd behaviour do not allow us to have a general argument for the nature of the bifurcation for any $k_f$.
On the other hand, we notice that even for the spatially extended system with $k_f \gg 1$ the mode $k_y = 1$ is always excited. 
Note, however, that the largest amplitude \sk{growing} mode for $k_f=3$ is not the $k_y = 1$ and it differs from the other cases. The $k_f = 3$ is also the only odd forcing case that gives a Hopf bifurcation, see Table \ref{tbl:kfLy}. 

It is commonly believed that the influence of the side walls on the instability should decrease when $k_f$ becomes large such that the behaviour predicted for unbounded Kolmogorov flows~\cite{thess92} should be recovered, i.e. a pitchfork bifurcation. This is not the case. The side walls, however distant, affect the nature of the bifurcation depending on the odd (respectively even) number of half-wavelengths of the base flow in the channel. This behaviour traces back to the large scale flow with $k_y = 1$ that is generated at the instability onset even for  $k_f$ large. A similar mechanism where distant side walls affect the nature of a bifurcation, has been described in the context of thermal convection~\cite{hirschberg1997}.

\begin{figure}[!ht]
  \vspace{0.3cm}
   \includegraphics[width=0.5\textwidth]{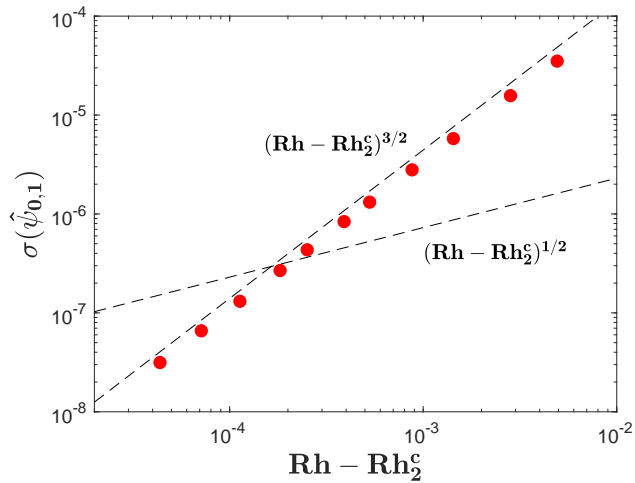}
 \caption{(Color online) The standard deviation of the largest scale mode $\widehat{\psi}_{0,1}$ as a function of the distance to the threshold $Rh - Rh^c_2$. The dashed lines indicate the scalings $(Rh - Rh_2^c)^{3/2}$ and $(Rh - Rh_2^c)^{1/2}$ for comparison.}
 \label{fig:sec_inst}
 \end{figure}

Now, we return to the case of $k_f = 4$ where the oscillating flow obtained 
for $Rh > Rh_1^c \approx 0.593$ persists up to $Rh = Rh_2^c \approx 0.835$,
above which a Hopf bifurcation takes place 
and the largest scale mode of the system $\widehat{\psi}_{0,1}$ is excited. 
Fig. \ref{fig:sec_inst} shows the standard deviation of $\widehat{\psi}_{0,1}$, denoted by $\sigma ( \widehat{\psi}_{0,1} )$, as a function of the distance to the threshold $Rh - Rh_2^c$ found from the DNS. The scaling we observe is $\sigma ( \widehat{\psi}_{0,1} ) \propto \left( Rh - Rh_2^c \right)^{3/2}$.
This is distinctively different from the standard $1/2$ power-law scaling one expects for the saturated amplitude of the \sk{growing mode} in the case of a supercritical Hopf bifurcation. 
This scaling results from the nonlinear excitation of the $\widehat{\psi}_{0,1}$ mode by the \sk{growing modes} that bifurcate at $Rh =Rh_2^c$. A low-dimensional model is presented in section \ref{sec:psi01model} to explain this behaviour.

\section{Low-dimensional dynamical systems \label{sec:12bif_the}}

\subsection{Four-mode model for the degenerate Hopf bifurcation \label{sec:psi31model}}

We first comment on the effect of the lateral confinement of the flow on the nature of the primary bifurcation. In the case of doubly periodic boundary conditions, 
Eq. \eqref{eq:NS} has the following symmetries among others; the translational symmetry along $x$, and the mirror symmetry with respect to $x$ and shift along $y$,
%
\begin{align}
&\psi(x,y,t) \to \psi(x + x_0,y,t), 
\label{translation} \\
&\psi(x,y,t) \to -\psi(-x,y + \frac{\pi}{k_f},t).
\label{mirrorandshift}
\end{align}
The perturbation that grows above threshold is to leading order of the form
\begin{equation}
\widetilde{ \psi } (x, y, t) = A(t) \phi (y) \exp i k_c x + c.c.,
\end{equation} 
where $A$ is the complex amplitude of the neutral mode of wavenumber $k_c$, $\phi (y)$ represents its dependence on $y$ and ``c.c." stands for the complex conjugate. 
Following \cite{fauve98} we expand $\dot A$ as a power series in $A$ and $A^{*}$ and we consider symmetry \eqref{translation} to constrain the form of the amplitude equation to
\begin{equation}
\dot A = \alpha A - \beta A^2  A^{*}, \label{eqn:amp_eq_simp}
\end{equation}
where $\alpha = \alpha_r + i \alpha_i$, $\beta = \beta_r + i \beta_i$ are some complex coefficients and the asterix ${\,}^{*}$ denotes the complex conjugate.
When $\phi(y)$ of the neutral mode is \sk{either symmetric or anti-symmetric under the shift $y \to y + \pi/k_f$, the transformation $x \rightarrow -x$ amounts to $A \rightarrow \pm A^{*}$}. The symmetry \eqref{mirrorandshift} 
implies that the amplitude equation should be invariant 
under the transformation $A \rightarrow A^{*}$. Taking the complex conjugate of the resulting equation implies $\alpha_i = \beta_i =0$. Therefore, the bifurcation is stationary as observed in the case of doubly periodic boundary conditions \cite{nepomniashchii1976,sivashinsky1985weak,lucaskerswell14}.

In the case of stress-free lateral boundary conditions, the shift along the $y$-axis is no longer possible 
and hence the symmetry \eqref{mirrorandshift} does not exist.
The imaginary parts of the coefficients of the amplitude equation are not constrained to vanish and we therefore expect a Hopf bifurcation as observed when only one wave length of the base flow fits in the channel \cite{kolesnikov1985,thess92,chen2002}. In the present case of 
$k_f = 4$, linear stability analysis shows that $\alpha_i = 0$, 
while $\beta_i \neq 0$, 
which explains the observed behaviour of the oscillation frequency. For a supercritical Hopf bifurcation ($\beta_r > 0$), the stationary amplitude squared of the limit cycle is given by $A A^{*} = \alpha_r/\beta_r$ and its phase is $\theta = -(\alpha_r \beta_i / \beta_r) t$. This gives a frequency of the limit cycle proportional to the distance to the threshold. Even so, there is no general argument to show that $\alpha_i = 0$ and it seems that this depends on the wavenumber $k_c$ \cite{chenprice05}. 

Note, however, that with stress-free boundary conditions, there are other symmetries that depend of the forcing wave number $k_f$.
If $k_f$ is even, the problem has mirror symmetry with respect to the mid-plane of the channel $y = \pi/2$
\begin{equation}
\psi(x,y,t) \to -\psi(x,\pi - y,t).
\label{eq:midplanesym}
\end{equation}
If $k_f$ is odd, this symmetry does no longer exist but we can find an invariance of the flow 
under the transformations
\begin{equation}
\psi(x,y,t) \to \psi(-x,\pi - y,t).
\label{eq:mirrorsyms}
\end{equation}
For all even values of $k_f$ in Table \ref{tbl:kfLy}, we find that the neutral mode has a $\phi (y)$ that is invariant under the transformation $y \to \pi - y$. Using this property of the neutral mode, we find that the symmetry \eqref{eq:midplanesym} does not give any new constraint on the amplitude equation \eqref{eqn:amp_eq_simp} in addition to the one that results from translational invariance. Thus the coefficients $\alpha$ and $\beta$ in Eq. \eqref{eqn:amp_eq_simp} can in general be complex for even values of $k_f$. Except for $k_f = 2$, we find that all the other even $k_f$ values that we examined, have $\alpha_i = 0$ and a complex $\beta$, see Table \ref{tbl:kfLy}. Now, for all odd values of $k_f$ except $k_f = 3$, we find that $\phi(y)$ of the neutral mode has a real part that is symmetric about the mid-line ($y = \pi/2$) and an imaginary part that is anti-symmetric about the mid-line. Thus in these cases under the transformation $y \to \pi - y$ we have $\phi (y) \to \phi^{*} (y)$. Applying the symmetry \eqref{eq:mirrorsyms} amounts to $A \to A^{*}$ in Eq. \eqref{eqn:amp_eq_simp}, which enforces a stationary bifurcation with $\alpha_i = \beta_i = 0$. This is true for all $k_f$ odd values except $k_f = 3$ as mentioned in Table \ref{tbl:kfLy}. The neutral mode in the case of $k_f = 3$ does not have any symmetry about the mid-line thus Eq. \eqref{eq:mirrorsyms} is not applicable and $\alpha, \beta$ can be complex. Thus from symmetry arguments using the form of the neutral mode we see why $k_f = 3$ is the only odd case in Table \ref{tbl:kfLy} that undergoes a Hopf bifurcation. 

\begin{figure}[!ht]
  \begin{subfigure}{0.25\textwidth}
   \includegraphics[width=\textwidth]{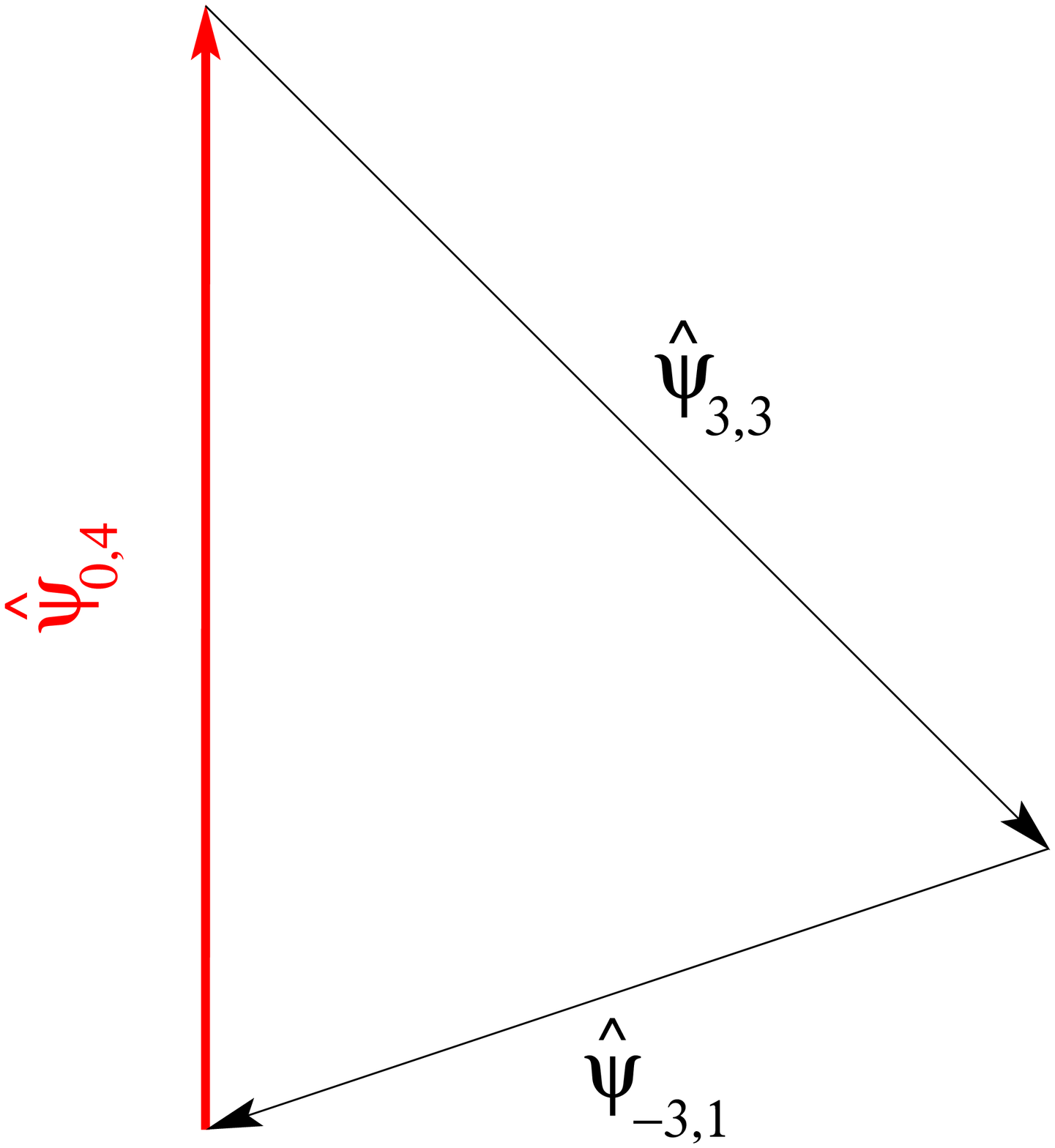}
   \caption{}
   \label{fig:triads1a}
  \end{subfigure}
  \hspace{2.0cm}
  \begin{subfigure}{0.23\textwidth}
  \includegraphics[width=\textwidth]{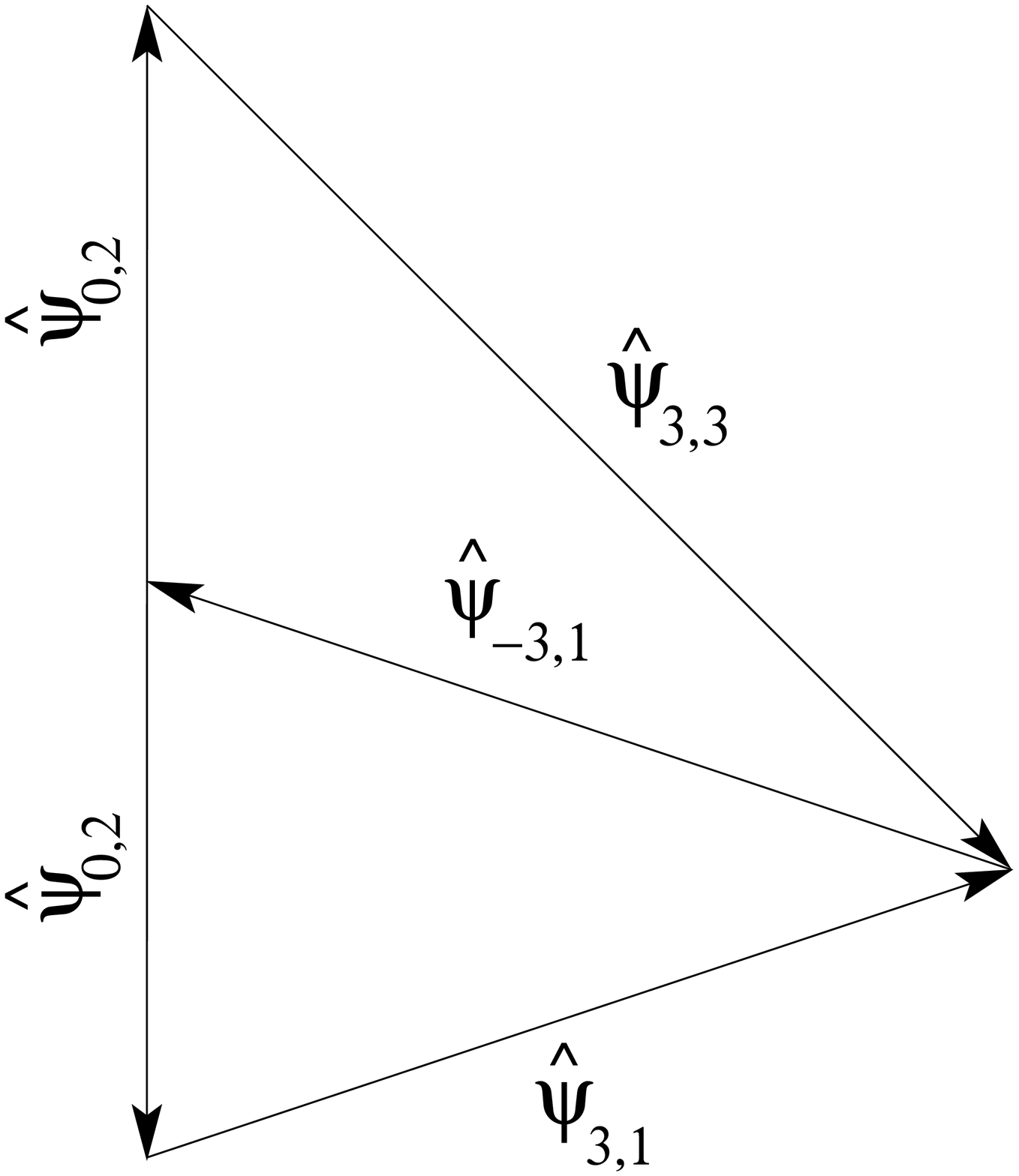}
     \caption{}
     \label{fig:triads1b}
  \end{subfigure}  
\caption{(Color online) Plots (a) and (b) show all the interacting triads in the reduced four-mode model. In plot (b) the mode $\widehat{\psi}_{\bf r} = \widehat{\psi}_{0,2}$ is repeated twice as it is excited by the modes $\widehat{\psi}^{*}_{\bf p} = \widehat{\psi}_{-3,1}$, $\widehat{\psi}^{*}_{\bf q} = \widehat{\psi}_{3,3}$ and is also responsible for the oscillation of the mode $\widehat{\psi}_{3,1}$. The red arrow indicates the base flow mode $\widehat{\psi}_{0,4}$, that becomes unstable at the first threshold $Rh_1^c$.  }
     \label{fig:triads1}
 \end{figure}

Now, we present a four-mode model to get a better qualitative understanding of this degenerate Hopf bifurcation for the case $k_f = 4$. To derive the governing equations we consider the Navier-Stokes equation in the Fourier-Sine basis form by substituting Eq. \eqref{eqn:normalmodes} into Eq. \eqref{eq:NS} to obtain 
\begin{equation}
d_t {\lhat\psi_{\bf k}} = 
 \sum_{{\bf p}, {\bf q}} A_{{\bf k},{\bf p},{\bf q}} \lhat\psi_{\bf p} \lhat\psi_{\bf q}
 - (Re^{-1} k^2 + Rh^{-1}) \lhat\psi_{\bf k} + f_0 \delta_{k_x,0} \, \delta_{k_y,4}
 \label{eq:NSmodes}
\end{equation}
with the interaction coefficients to be given by 
\begin{align}
A_{{\bf k},{\bf p},{\bf q}} = \frac{i}{2}(q^2 - p^2)k^{-2}\delta_{k_x,p_x + q_x}
[ (p_xq_y - p_yq_x) \delta_{k_y,p_y + q_y} 
+(p_x q_y + p_y q_x) (\delta_{k_y,p_y - q_y} - \delta_{k_y,q_y - p_y})],
 \label{eq:NSkernel}                                                       
\end{align}
where $\delta_{i,j}$ stands for the Kronecker delta. 
Consider a model with the base flow $\widehat{\psi}_{0,4}$, the two \sk{largest amplitude growing modes} $\widehat{\psi}_{3,1}, \widehat{\psi}_{-3,3}$ and a nonlinear mode $\widehat{\psi}_{0,2}$ excited by the \sk{two growing modes} denoted as,
\begin{equation}
\widehat{\psi}_{\bf k} = \widehat{\psi}_{0,4}, \quad 
\widehat{\psi}_{\bf p} =\widehat{\psi}_{3,1}, \quad
\widehat{\psi}_{\bf q} = \widehat{\psi}_{-3,3}, \quad
\widehat{\psi}_{\bf r} = \widehat{\psi}_{0,2}.  
\end{equation}
Using Eqs. \eqref{eq:NSmodes}-\eqref{eq:NSkernel}, we arrive at the following system of equations
\begin{align}
 d_t{\widehat{\psi}_{\bf k}} &+  \left( 16 Re^{-1} + Rh^{-1} \right) \widehat{\psi}_{\bf k} = - 3 i \left( \widehat{\psi}_{\bf p}^{*} \widehat{\psi}_{\bf q}^{*} - \widehat{\psi}_{\bf p} \widehat{\psi}_{\bf q} \right) + 1, \label{eqn:fourmode_1} \\
 d_t{\widehat{\psi}_{\bf p}} &+  \left( 10 Re^{-1} + Rh^{-1} \right) \widehat{\psi}_{\bf p} =  \frac{6}{5} i \widehat{\psi}_{\bf k} \widehat{\psi}_{\bf q}^{*} - \frac{21}{5} i \widehat{\psi}_{\bf r} \widehat{\psi}_{\bf q}^{*} - \frac{9}{5} i  \widehat{\psi}_{\bf p} \widehat{\psi}_{\bf r}, \label{eqn:fourmode_2} \\
 d_t{\widehat{\psi}_{\bf q}} &+ \left( 18 Re^{-1} + Rh^{-1} \right) \widehat{\psi}_{\bf q} = 2 i \widehat{\psi}_{\bf k} \widehat{\psi}_{\bf p}^{*} + i \widehat{\psi}_{\bf r} \widehat{\psi}_{\bf p}^{*},  \label{eqn:fourmode_3} \\
 d_t{\widehat{\psi}_{\bf r}}  &+   \left(   4 Re^{-1} + Rh^{-1} \right) \widehat{\psi}_{\bf r} =  6 i \left( \widehat{\psi}_{\bf p}^{*} \widehat{\psi}_{\bf q}^{*} - \widehat{\psi}_{\bf p} \widehat{\psi}_{\bf q} \right).  
\label{eqn:fourmode_4}
\end{align}
The triads that can be constructed from this set of modes are shown in Fig. \ref{fig:triads1} and we will discuss their dynamics in what follows.
Here, 
the constant term on the right of Eq. \eqref{eqn:fourmode_1} is the forcing term which only acts on the mode $\widehat{\psi}_{\bf k}$. The base flow is given by the balance between the forcing and the dissipation in \eqref{eqn:fourmode_1}, which gives $\widehat{\psi}_{\bf k} = \psi_0 = 1/\left( 16 Re^{-1} + Rh^{-1} \right)$. The instability is found by linearising the above set of equations around the base flow $\psi_0$. In the linearised system, we see that only $\widehat{\psi}_{\bf p}, \widehat{\psi}_{\bf q}$ are coupled and the mode $\widehat{\psi}_{\bf r}$ is not coupled to the base flow. 
Then, the effective linearised system can be written as
\begin{align}
d_t{\widehat{\psi}_{\bf p}} + \left( 10 Re^{-1} + Rh^{-1} \right) \widehat{\psi}_{\bf p} & =    
\frac{12 i}{10} {\psi}_0 \widehat{\psi}_{\bf q}^{*}, \label{eqn:linearized_1} \\
d_t{\widehat{\psi}_{\bf q}^{*}} + \left( 18 Re^{-1} + Rh^{-1} \right) \widehat{\psi}_{\bf q}^{*} & = 
- 2 i \psi_0 \widehat{\psi}_{\bf p}. \label{eqn:linearized_2}
\end{align}
The linear stability threshold is found to be at \sk{$Rh^{c}_1 \approx 0.814$} 
and the Jacobian of Eqs. \eqref{eqn:linearized_1}, \eqref{eqn:linearized_2} gives two eigenmodes. Both eigenvalues are purely real leading to exponential growth or decay with no oscillations. 
We denote the amplitudes of the \sk{decaying} and the \sk{growing} eigenmode as $P_1(t)$ and $P_2(t)$, respectively. The positive eigenvalue $\lambda_2$ that corresponds to the \sk{growing eigenmode} $P_2$, scales linearly with the distance to the threshold 
\begin{equation}
 \lambda_2 \propto (Rh - Rh_1^c), 
 \label{eqn:lambda_scal}
\end{equation}
This scaling is true only close to the threshold. Its exact expression is given in Appendix \ref{App1}.

Now we choose to solve the nonlinear model with only the linearly excited modes, i.e. we consider only the 
triad $(\widehat{\psi}_{\bf k}, \widehat{\psi}_{\bf p}, \widehat{\psi}_{\bf q})$ given by Eqs. \eqref{eqn:fourmode_1}, \eqref{eqn:fourmode_2}, \eqref{eqn:fourmode_3} (see Fig. \ref{fig:triads1a}) and with $\widehat{\psi}_{\bf r} = 0$. 
The modes 
$\widehat{\psi}_{\bf p}$, $\widehat{\psi}_{\bf q}$ are linearly excited by the instability of the base flow and saturate by modifying the amplitude of the mode $\widehat{\psi}_{\bf k}$.
We then solve the full system of equations \eqref{eqn:fourmode_1}, \eqref{eqn:fourmode_2}, \eqref{eqn:fourmode_3} by focusing on the evolution of the \sk{growing} eigenmode $P_2$. The resulting amplitude equation is
\begin{equation}
d_t{P_2} = \lambda_2 P_2 - \left( \frac{\frac{36}{5} \zeta_2 \beta_1 + 12 \zeta_2^2 \beta_3}{16 Re^{-1} + Rh^{-1}} \right) | P_2 |^2 P_2. 
\label{eqn:threemode}
\end{equation}
%
Details for its derivation and the 
expressions of the real coefficients 
$\zeta_2, \beta_1, \beta_3$ can be found in Appendix \ref{App1}. 
Thus, the amplitude equation for the three mode model $\widehat{\psi}_{\bf k}$, $\widehat{\psi}_{\bf p}$, $\widehat{\psi}_{\bf q}$ clearly gives rise to a stationary bifurcation with 
the scaling of the amplitude of the \sk{growing mode} to be $|P_2| \propto (Rh - Rh_c)^{1/2}$ obtained from Eq. \eqref{eqn:lambda_scal}.

If we now consider Eq. \eqref{eqn:fourmode_4}, we see that the mode $\widehat{\psi}_{\bf r}$ is nonlinearly excited by the modes $\widehat{\psi}_{\bf p}, \widehat{\psi}_{\bf q}$ (see Fig. \ref{fig:triads1b}). This nonlinear excitation arises due to the transfer of energy from modes $(p_x, p_y), (q_x, q_y)$ to both $(-p_x - q_x,p_y+q_y)$ and $(-p_x - q_x, |p_y - q_y|)$ in the Fourier-Sine basis. Taking into account all the four modes we can get to the following amplitude equation
\begin{align}
d_t{P_2} = \lambda_2  P_2 
- \left[ \left( \frac{\frac{36}{5} \zeta_2 \beta_1 + 12 \zeta_2^2 \beta_3}{16 Re^{-1} + Rh^{-1}} \right) 
       + \left( \frac{\frac{252}{5} \zeta_2 \beta_1 - 12 \zeta_2^2 \beta_3}{4 Re^{-1} + Rh^{-1}} \right) \right] |P_2|^2 P_2 
- i \left( \frac{108}{5} \frac{\zeta_2^2 \beta_1}{4 Re^{-1} + Rh^{-1}} \right) |P_2|^2 P_2,
\label{eqn:fourmode}
\end{align}
where the expressions for $\lambda_2, \beta_1, \beta_3, \zeta_2$ are given in Appendix \ref{App1}. This amplitude equation is very similar to Eq. \eqref{eqn:threemode}, the amplitude equation for the three mode model, except for the presence of the complex coefficient in the final term which arises from the 
existence of $\widehat{\psi}_{\bf r}$. This new term leads to oscillatory solutions of the form $P_2 (t) = | P_2 | \exp \left( i \omega_f t \right)$. By substituting this solution into Eq. \eqref{eqn:fourmode} we get
\begin{align}
|P_2|^2 &= \lambda_2 
\left[ \left( \frac{\frac{36}{5} \zeta_2 \beta_1 + 12 \zeta_2^2 \beta_3}{16 Re^{-1} + Rh^{-1}} \right) 
       + \left( \frac{\frac{252}{5} \zeta_2 \beta_1 - 12 \zeta_2^2 \beta_3}{4 Re^{-1} + Rh^{-1}} \right) \right]^{-1}, \\
\omega_f &= - \frac{108}{5} \frac{\zeta_2^2 \beta_1}{4 Re^{-1} + Rh^{-1}} |P_2|^2.
\end{align}
Using Eq. \eqref{eqn:lambda_scal} we find the amplitude to scale similar to the three mode model, i.e. $|P_2| \propto \left( Rh - Rh_c \right)^{1/2}$
and the oscillation frequency to scale linearly with the distance to the threshold $\omega_f \propto \left( Rh - Rh_c \right)$. Thus, the minimal four-mode model reproduces the degenerate Hopf bifurcation and the observed scalings of the DNS results. 
The value we obtain for the threshold does not agree quantitatively with the DNS. This is because in the DNS many modes are non-zero in contrast to our minimal model which only considers the four modes with the largest amplitude in the full system. By adding more modes and following the method presented above we can approach the values of the threshold and the oscillation frequency obtained in the DNS.

\subsection{Large scale bifurcation model \label{sec:psi01model}}

Here we present a model to explain the exponent $3/2$ for the largest scale mode in the system $\widehat{\psi}_{0,1}$. This mode is directly excited by the nonlinear perturbations that grow after the second Hopf bifurcation. We present here a model with eight modes that is sufficient to capture the different scalings needed to explain the exponent $3/2$. The base flow over which the second instability develops involves multiple modes. We construct a reduced model using the following set of modes,
\begin{equation}
\widehat{\psi}_{\bf a} = \widehat{\psi}_{-3,1}, \quad 
\widehat{\psi}_{\bf b} = \widehat{\psi}_{-3,3}, \quad
\widehat{\psi}_{\bf c} = \widehat{\psi}_{1,4}, \quad
\widehat{\psi}_{\bf d} = \widehat{\psi}_{2,5}, \quad
\widehat{\psi}_{\bf e} = \widehat{\psi}_{2,1},  \quad
\widehat{\psi}_{\bf f} = \widehat{\psi}_{-5,2}, \quad
\widehat{\psi}_{\bf g} = \widehat{\psi}_{-1,3}, \quad
\widehat{\psi}_{\bf h} = \widehat{\psi}_{0,1}. 
\end{equation}

\begin{figure}[!ht]
  \begin{subfigure}{0.245\textwidth}
   \includegraphics[height=5.0cm]{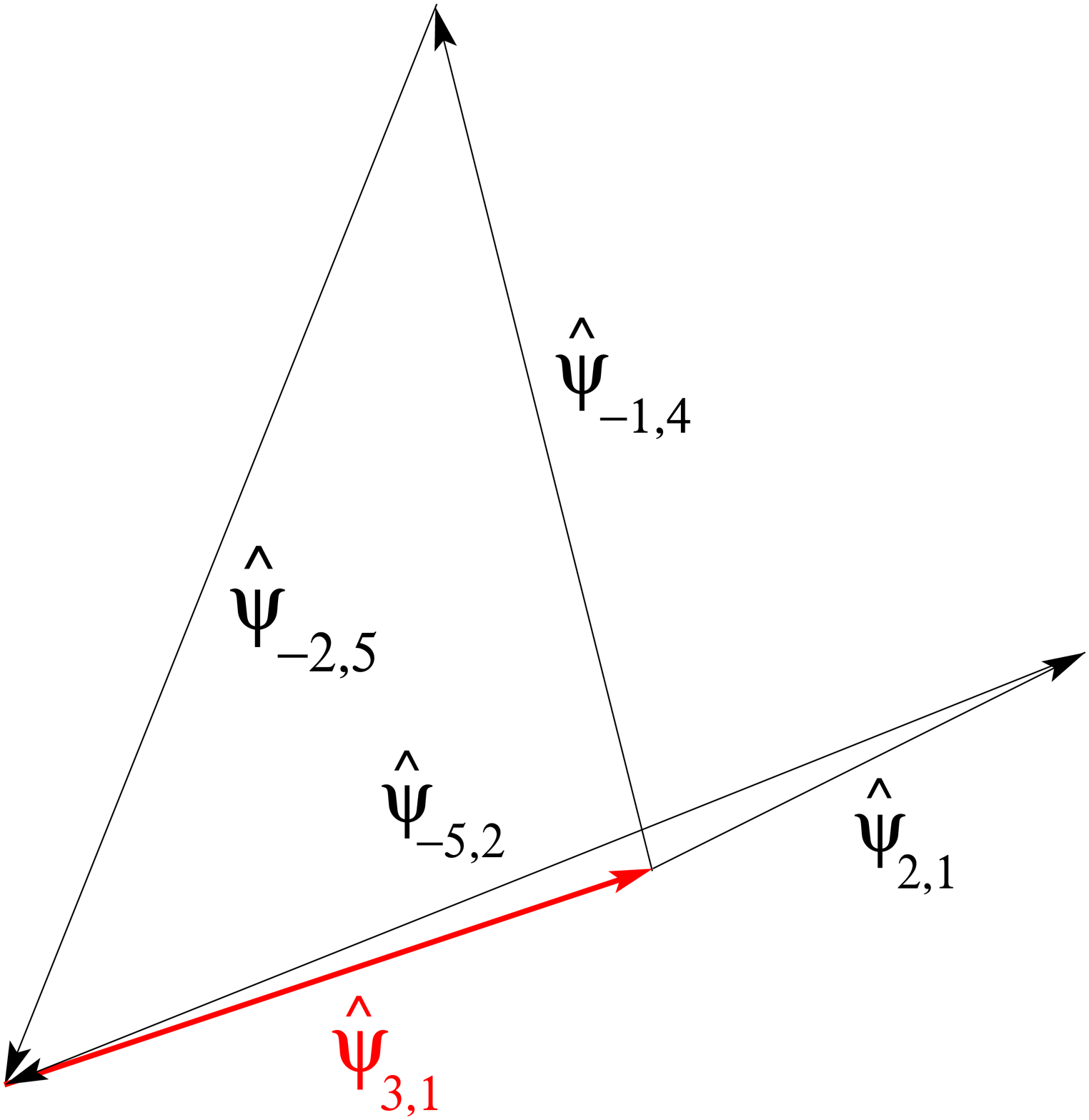}
   \caption{}
   \label{fig:triads2a}
  \end{subfigure}
  \begin{subfigure}{0.245\textwidth}
  \includegraphics[height=5.0cm]{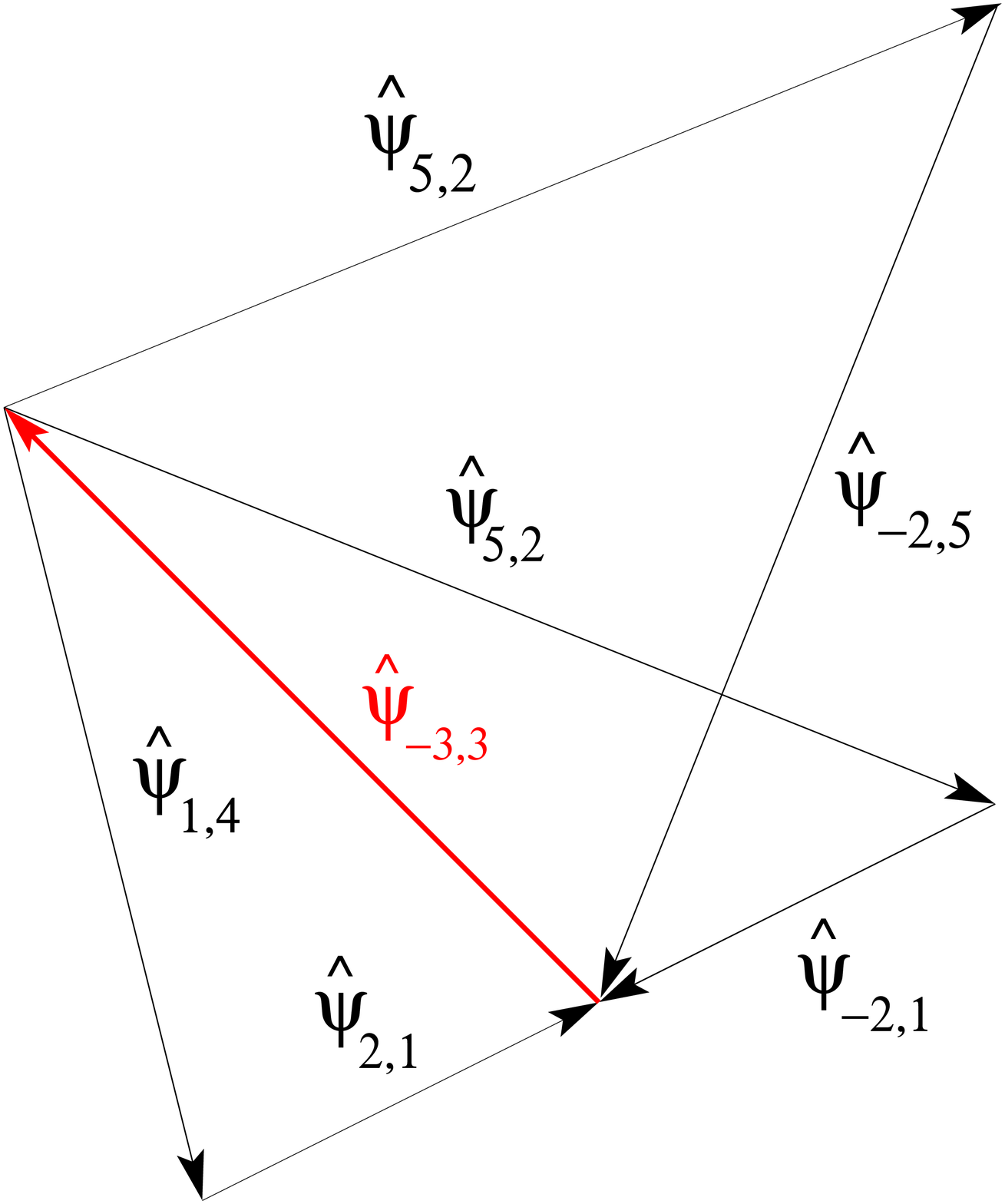}
     \caption{}
     \label{fig:triads2b}
  \end{subfigure}
  \begin{subfigure}{0.245\textwidth}
  \includegraphics[height=5.0cm]{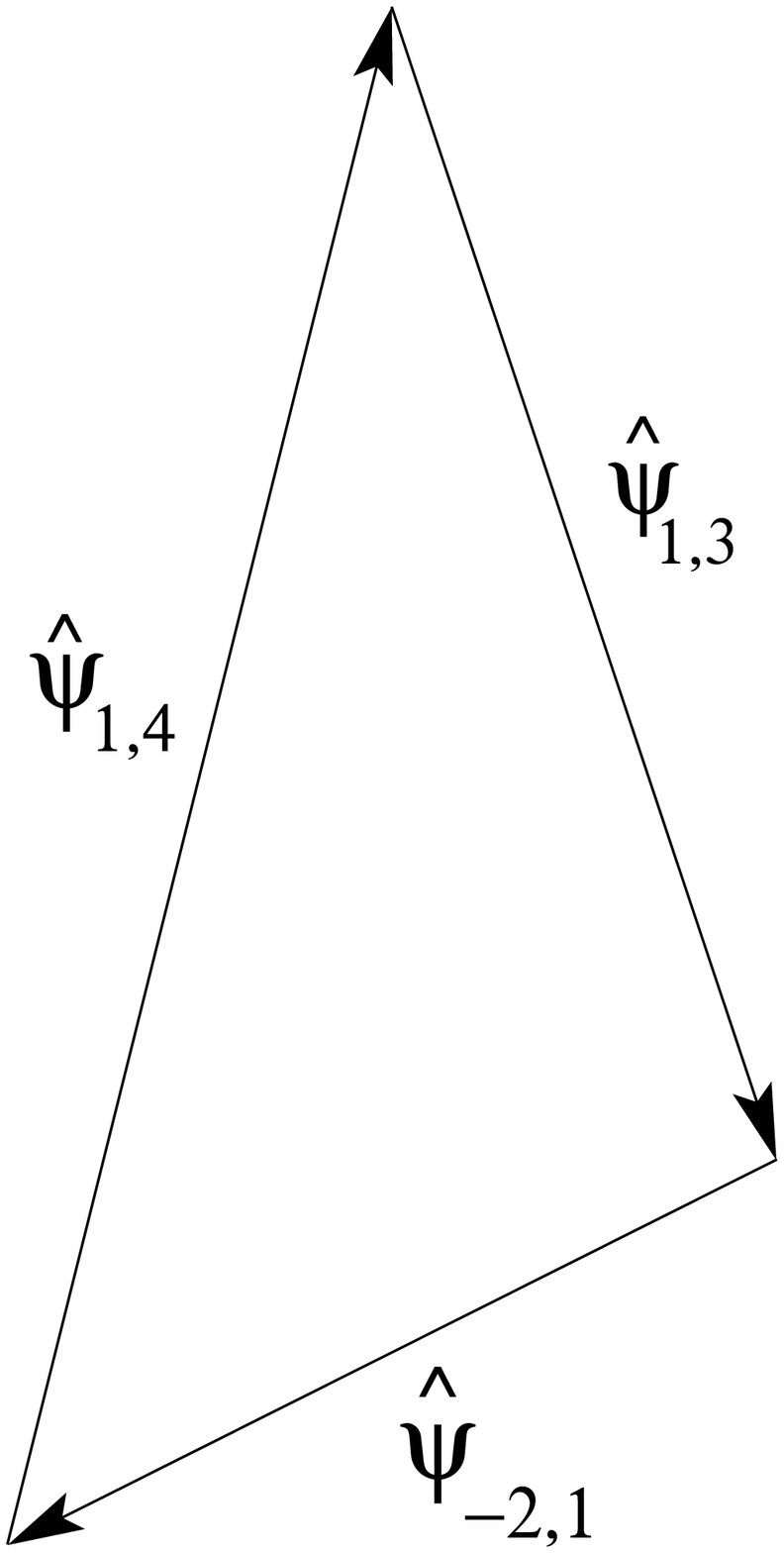}
     \caption{}
     \label{fig:triads2c}
  \end{subfigure}
  \begin{subfigure}{0.245\textwidth}
   \includegraphics[height=5.0cm]{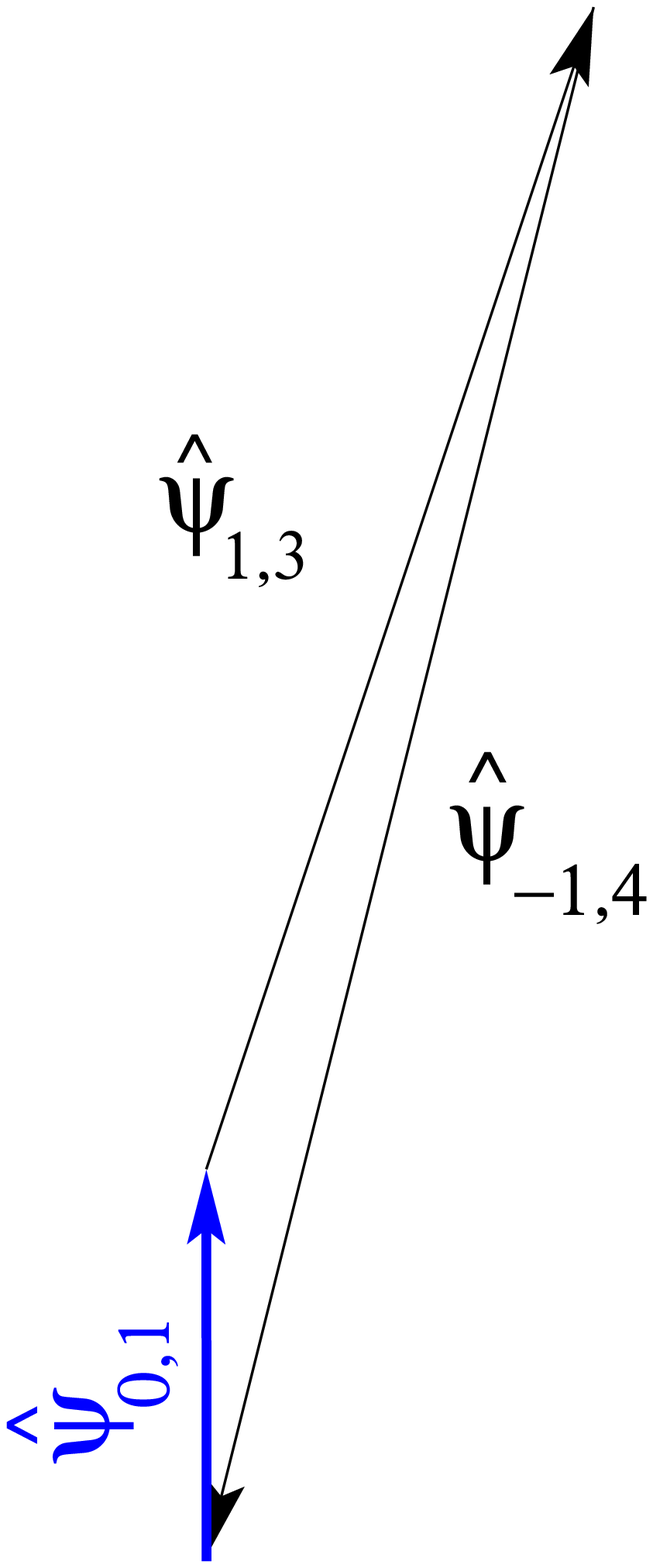}
      \caption{}
      \label{fig:triads2d}
  \end{subfigure}
 \caption{(Color online) Plots (a), (b), (c) and (d) show all the interacting triads of the reduced eight-mode model. The red arrows indicate the modes that become unstable at the second threshold $Rh_2^c$ and the blue arrow indicates the large scale mode $\widehat{\psi}_{0,1}$.}
     \label{fig:triads2}
 \end{figure}

These modes are chosen because they have the largest amplitudes in the DNS. Moreover, we tested that if any of the modes are put to zero, then $\widehat{\psi}_{0,1}$ is not excited or it has a much lower amplitude. This demonstrates how vital these  modes are to the excitation of the large scale mode. Below the second instability all the aforementioned modes 
have zero amplitude apart from the modes $\widehat{\psi}_{\bf a} = \widehat{\psi}_{-3,1}, \widehat{\psi}_{\bf b} = \widehat{\psi}_{-3,3}$, which are already excited at the first instability.  
Above the threshold value of $Rh_2^c$, the modes $\widehat{\psi}_{\bf a} = \widehat{\psi}_{-3, 1}, \widehat{\psi}_{\bf b} = \widehat{\psi}_{-3, 3}$ become linearly unstable and give rise to the modes $\widehat{\psi}_{\bf c}, \widehat{\psi}_{\bf d}, \widehat{\psi}_{\bf e}, \widehat{\psi}_{\bf f}$. The triadic interactions are shown in Figs. \ref{fig:triads2a} and \ref{fig:triads2b}.
The governing equations for the unstable modes $\widehat{\psi}_{\bf a}, \widehat{\psi}_{\bf b}$ and the linearly excited modes $\widehat{\psi}_{\bf c}, \widehat{\psi}_{\bf d}, \widehat{\psi}_{\bf e}, \widehat{\psi}_{\bf f}$ are,
\begin{align}
d_t{\widehat{\psi}_{\bf a}} &+ (10 Re^{-1} + Rh^{-1}) \widehat{\psi}_{\bf a} = 
  i\frac{39}{5} \widehat{\psi}_{\bf c}^{*} \widehat{\psi}_{\bf d}^{*} 
+ i\frac{6}{5} \widehat{\psi}_{\bf e} \widehat{\psi}_{\bf f} + f_1, \\
d_t{\widehat{\psi}_{\bf b}} & + (18 Re^{-1} + Rh^{-1}) \widehat{\psi}_{\bf b} = 3 i \widehat{\psi}_{\bf c}^{*} \widehat{\psi}_{\bf e}^{*} + 6 i \widehat{\psi}_{\bf e} \widehat{\psi}_{\bf f} + f_2, \\
d_t{\widehat{\psi}_{\bf c}} &+ \left( 17 Re^{-1} + Rh^{-1} \right) \widehat{\psi}_{\bf c} = -i\frac{247}{34} \widehat{\psi}_{\bf d}^{*} \widehat{\psi}_{\bf a}^{*} - i \frac{117}{34} \widehat{\psi}_{\bf e}^{*} \widehat{\psi}_{\bf b}^{*} + i \frac{35}{34} \widehat{\psi}_{\bf e} \widehat{\psi}_{\bf g} - i \frac{9}{34} \widehat{\psi}_{\bf g}^{*} \widehat{\psi}_{\bf h}, \\
d_t{\widehat{\psi}_{\bf d}} &+ \left( 29 Re^{-1} + Rh^{-1} \right) \widehat{\psi}_{\bf  d} = i\frac{91}{58} \widehat{\psi}_{\bf c}^{*} \widehat{\psi}_{\bf a}^{*} - i \frac{231}{58} \widehat{\psi}_{\bf f}^{*} \widehat{\psi}_{\bf b}, \\
d_t{\widehat{\psi}_{\bf e}} &+ \left( 5 Re^{-1} + Rh^{-1} \right) \widehat{\psi}_{\bf e} = i\frac{19}{10} \widehat{\psi}_{\bf f}^{*} \widehat{\psi}_{\bf a} + i \frac{9}{10} \widehat{\psi}_{\bf c}^{*} \widehat{\psi}_{\bf b}^{*} + i \frac{99}{10} \widehat{\psi}_{\bf f}^{*} \widehat{\psi}_{\bf b} - i \frac{49}{10} \widehat{\psi}_{\bf c} \widehat{\psi}_{\bf g}^{*}, \\
d_t{\widehat{\psi}_{\bf f}} &+ \left( 29 Re^{-1} + Rh^{-1} \right) \widehat{\psi}_{\bf  f} = i\frac{5}{58} \widehat{\psi}_{\bf e}^{*} \widehat{\psi}_{\bf a} + i \frac{231}{58} \widehat{\psi}_{\bf d}^{*} \widehat{\psi}_{\bf b} + i \frac{117}{58} \widehat{\psi}_{\bf e}^{*} \widehat{\psi}_{\bf b}.
\end{align}
The terms $f_1$, $f_2$ denote the forcing due to the first instability and contain the interaction terms with the modes presented in the previous section. At saturation the amplitudes of the modes $|\widehat{\psi}_{\bf c}|$, $|\widehat{\psi}_{\bf d}|$, $|\widehat{\psi}_{\bf e}|$, $|\widehat{\psi}_{\bf f}|$ scale like $(Rh - Rh_2^c)^{1/2}$. Next, we consider the triadic interaction between the modes $(\widehat{\psi}_{\bf c}, \widehat{\psi}_{\bf e}^*, \widehat{\psi}_{\bf g}^*)$ (see Fig. \ref{fig:triads2c}). 
In this triad, the mode $\widehat{\psi}_{\bf g}^* =  \widehat{\psi}_{1,3}$ is excited by the nonlinear interaction between the two linearly \sk{excited} modes $\widehat{\psi}_{\bf c} = \widehat{\psi}_{1,4}$ and $\widehat{\psi}_{\bf e}^* = \widehat{\psi}_{-2,1}$. The governing equation for $\widehat{\psi}_{\bf g}$  is given by
\begin{align}
d_t{\widehat{\psi}_{\bf g}} + \left( 10 Re^{-1} + Rh^{-1} \right) \widehat{\psi}_{\bf g} = i \frac{21}{5} \widehat{\psi}_{\bf c} \widehat{\psi}_{\bf e}^{*} + i \frac{4}{5} \widehat{\psi}_{\bf h} \widehat{\psi}_{\bf c}^{*}, 
\label{eqn:nonlin_forc1}
\end{align}
which at saturation gives rise to the scaling $|\widehat{\psi}_{\bf g}| \propto (Rh - Rh_2^c)$ for the amplitude. Then, we consider the triad $(\widehat{\psi}_{\bf c}^*, \widehat{\psi}_{\bf g}^*, \widehat{\psi}_{\bf h})$, where the mode $\widehat{\psi}_{\bf h} = \widehat{\psi}_{0,1}$ is excited by the nonlinear interaction between the linearly \sk{excited} mode $\widehat{\psi}_{\bf c}^* = \widehat{\psi}_{-1,4}$ and the nonlinearly excited mode $\widehat{\psi}_{\bf g}^* = \widehat{\psi}_{1,3}$ (see Fig. \ref{fig:triads2d}). The governing equation for $\widehat{\psi}_{\bf h}$ is given by
\begin{align}
d_t{\widehat{\psi}_{\bf h}} + \left( Re^{-1} + Rh^{-1} \right) \widehat{\psi}_{\bf h} = i\frac{7}{2} \left( \widehat{\psi}_{\bf c} \widehat{\psi}_{\bf g} - \widehat{\psi}_{\bf c}^{*} \widehat{\psi}_{\bf g}^{*} \right).
\end{align}
Thus, at saturation the mode scales like $|\widehat{\psi}_{\bf h}| = |\widehat{\psi}_{0,1}| \propto \left( Rh - Rh_2^c \right)^{3/2}$. 
To sum up, this reduced order model captures all the necessary scalings that were observed in the DNS for the second Hopf bifurcation. 

\section{Conclusion \label{sec:conclus}}

We have studied the primary bifurcations of a forced Kolmogorov flow in a channel with free-slip boundary conditions in the lateral direction and periodic boundary conditions in the longitudinal direction, our aim being to mimic a parallel flow. 

Unlike the doubly periodic Kolmogorov flow, where the first bifurcation is stationary, a simple symmetry argument shows why we can expect it to become a Hopf bifurcation in the case of a laterally confined flow. This qualitative change is observed in direct numerical simulations. However, at the threshold of the first instability $Rh_1^c$ we find a new type of bifurcation where both the amplitude and the oscillation frequency of the \sk{growing} mode are zero. As we move away from the threshold the amplitude scales with an exponent $1/2$ and the oscillation frequency with an exponent $1$ of the distance to the threshold. Although the linear stability analysis shows that the growth rate is real as for a stationary bifurcation, the leading order nonlinear term in the amplitude equation has a complex coefficient as in the case of a Hopf bifurcation. This explains the scalings observed in direct numerical simulations. We call this bifurcation a degenerate Hopf bifurcation. We have not found a general argument to show that the linear growth rate is real, i.e. some principle of exchange of stability \cite{chandrasekhar61} for $k_f \geq 4$. It is unlikely that such a principle exists. None has been found in the case of other parallel flows~\cite{drazin1981}.
One scenario that merits further studies is that 
the nature of the bifurcation also depends on the instability wave number $k_x$ in the case of a channel with a large aspect ratio~\cite{chen20}.
Using a truncated model, we derived an amplitude equation with the required properties. This reduced model displays the scaling observed in the DNS. We expect that degenerate Hopf bifurcations can also occur in the case of other parallel flows.

A secondary instability occurs when $Rh=Rh_2^c$ and corresponds to a Hopf bifurcation. This leads to the generation of the largest scale mode in the system $\widehat{\psi}_{0,1}$, i.e. a large scale shear flow with half a wave length in the spanwise direction and no dependence on the streamwise direction. Its amplitude displays a surprising scaling $\left(Rh - Rh_2^c \right)^{3/2}$. Using a truncated model, we find that the large scale shear is not a bifurcating mode for $Rh=Rh_2^c$ but is nonlinearly excited by the bifurcating modes, which explains the observed scaling. It is surprising that these odd scaling laws at first sight are not reported more often in experiments where it is difficult to determine if the measured quantity is proportional to the amplitude of the bifurcating modes or to their harmonics.

\section*{Acknowledgements}

The authors would like to thank 
A. Alexakis, J. Chapman, P. J. Ioannou and T. Mullin for useful discussions. 


\appendix

\section{Derivation of the amplitude equations} \label{App1}

Here we provide details for the derivation of the amplitude equations \eqref{eqn:threemode}, \eqref{eqn:fourmode} starting from the governing equations Eqs. \eqref{eqn:fourmode_1} - \eqref{eqn:fourmode_4}. 
We start with the linearized equations \eqref{eqn:linearized_1}, \eqref{eqn:linearized_2} written in the matrix form, $d_t {\bm \Psi} = \mathcal{A} {\bm \Psi}$, viz.
\begin{align}
d_t
\begin{bmatrix}
          {\widehat{\psi}_{\bf p}} \\
          {\widehat{\psi}_{\bf q}^{*}}
         \end{bmatrix} & =  \begin{bmatrix}
         - \left( 10 Re^{-1} + Rh^{-1} \right) & i\frac{12}{10} \psi_0 \\
         - i 2 \psi_0 & - \left( 18 Re^{-1} + Rh^{-1} \right) 
         \end{bmatrix} 
         \begin{bmatrix}
          \widehat{\psi}_{\bf p} \\
          \widehat{\psi}_{\bf q}^{*}
         \end{bmatrix}
         \label{eqn:matrix_eqn}
\end{align}
The solution to $\det(\mathcal{A} - \lambda \mathcal{I}) = 0$ gives the following two eigenvalues
\begin{align}
\lambda_1 & =  -\frac{2 \sqrt{20 Re^{-2} (Rh^{-1} +16 Re^{-1} )^2+3}}{\sqrt{5} (Rh^{-1} +16 Re^{-1} )}-Rh^{-1} -14 Re^{-1}, \\
\lambda_2 & = \frac{2 \sqrt{20 Re^{-2} (Rh^{-1} +16 Re^{-1} )^2+3}}{\sqrt{5} (Rh^{-1} +16 Re^{-1} )}-Rh^{-1} -14 Re^{-1}.
\end{align}
We see that the eigenvalue $\lambda_1$ is always negative and the threshold of the instability is found by putting $\lambda_2 = 0$, which gives the threshold $Rh^{-1}_c \approx 1.229$ for $Re = 1000$. The associated eigenvectors denoted as ${\bf V}_1$, ${\bf V}_2$ are given by
\begin{align}
{\bf V_1} = \begin{bmatrix}
          -\frac{1}{5} i \left(\sqrt{5} \sqrt{20 Rh^{-2} Re^{-2} + 640 Rh^{-1}  Re^{-3} + 5120 Re^{-4} + 3} - 10 Rh^{-1} Re^{-1} - 160 Re^{-2} \right) \\ 
          1
      \end{bmatrix}, \\ \vspace{0.5cm}
{\bf V_2} = \begin{bmatrix} 
          \frac{1}{5} i \left(\sqrt{5} \sqrt{20 Rh^{-2} Re^{-2} + 640 Rh^{-1}  Re^{-3} + 5120 Re^{-4} + 3}+10 Rh^{-1} Re^{-1} + 160 Re^{-2} \right) \\
          1
      \end{bmatrix}. 
\end{align}
We then express the variables ${\bm \Psi}(t)$ as a linear combination of the two eigenvectors with amplitudes $P_1 (t), P_2 (t)$,
\begin{align}
{\bm \Psi} (t) = P_1(t) {\bf V}_1 + P_2(t) {\bf V}_2 = \mathcal{V} {\bf P}, 
\label{eqn:relation_modes_app}
\end{align} 
where $\mathcal{V} = [{\bf V}_1 \,\, {\bf V}_2]$ denotes the eigenvector matrix and 
${\bf P} = 
\begin{bmatrix}
P_1 \\ P_2
\end{bmatrix}$ 
denotes the amplitude vector.


To get the nonlinear system of equations in terms of the eigenvectors, we start with
\begin{align}
d_t{\bm \Psi} & = \mathcal{A} {\bm \Psi} + \begin{bmatrix}
          -i\frac{6}{5} \widetilde{\psi}_{\bf k} \widehat{\psi}_{\bf q}^{*} - i\frac{21}{5} \widehat{\psi}_{\bf r} \widehat{\psi}_{\bf q}^{*} - i\frac{18}{5} \widehat{\psi}_{\bf p} \widehat{\psi}_{\bf r} \vspace{0.2cm} \\ 
          - 2 i \widetilde{\psi}_{\bf k} \widehat{\psi}_{\bf p}^{*} + i \widehat{\psi}_{\bf r} \widehat{\psi}_{\bf p}^{*}
         \end{bmatrix} \nonumber \\
& = \mathcal{A} {\bm \Psi} + {\bf N}, 
\end{align}
where ${\bf N}$ denotes the nonlinear terms and $\widetilde{\psi}_{\bf k}$ denotes the deviation of $\widehat{\psi}_{\bf k}$ from the base flow $\psi_0$ due to the nonlinearity, viz. $\widehat{\psi}_{\bf k} = \psi_0 - \widetilde{\psi}_{\bf k}$. Doing the eigendecomposition of the matrix $\mathcal{A}$, we get 
\begin{align}
 d_t{\bm \Psi} = \mathcal{V} \mathcal{D} \mathcal{V}^{-1} {\bm \Psi} + {\bf N}.
\end{align}
where $\mathcal{D}$ is a diagonal matrix whose diagonal entries are the eigenvalues $\lambda_1$ and $\lambda_2$. Taking $\mathcal{V}^{-1}$ on both sides leads to the equation,
\begin{align}
 d_t \left(\mathcal{V}^{-1} {\bm \Psi} \right) = \mathcal{D} \left( \mathcal{V}^{-1} {\bm \Psi} \right) + \mathcal{V}^{-1} {\bf N}.
\label{eqn:simp_nl_app}
\end{align}
From \eqref{eqn:relation_modes_app} we can write $\mathcal{V}^{-1} {\bm \Psi} = {\bf P}$, which gives
\begin{align}
 d_t {\bf P} = \mathcal{D} {\bf P} + \mathcal{V}^{-1} {\bf N}.
\label{eqn:ampl_eqn_app}
\end{align}
Since $\lambda_1 < 0$ and the nonlinear terms do not force $P_1(t)$, the amplitude $|P_1|$ goes to zero in the long time limit, implying that we can express $\widehat{\psi}_{\bf p}, \widehat{\psi}_{\bf q}$ in terms of $P_2 (t)$ only. 
The nonlinear 
vector ${\bf N}$ involves terms with $\widetilde{\psi}_{\bf k}$ 
that is non-zero above the first instability threshold 
and is modified by terms involving $\widehat{\psi}_{\bf p}, \widehat{\psi}_{\bf q}$ 
(see Eq. \eqref{eqn:fourmode_1}). 
The ${\bf N}$ vector also involves $\widehat{\psi}_{\bf r}$, which is excited by terms involving $\widehat{\psi}_{\bf p}, \widehat{\psi}_{\bf q}$ (see Eq. \eqref{eqn:fourmode_4}).
The nonlinear term needed to find $\widetilde{\psi}_{\bf k}$, and $\widehat{\psi}_{\bf r}$ is given by the expression $\widehat{\psi}_{\bf p}^* \widehat{\psi}_{\bf q}^* - \widehat{\psi}_{\bf p} \widehat{\psi}_{\bf q}$ (see Eqs. \eqref{eqn:fourmode_1}, \eqref{eqn:fourmode_4}). This can be written in terms of $P_2 (t)$ as, 
\begin{align}
\widehat{\psi}_{\bf p}^* \widehat{\psi}_{\bf q}^* - \widehat{\psi}_{\bf p} \widehat{\psi}_{\bf q} \rightarrow - 2 i \zeta_2 |P_2|^2,
\end{align}
where $\zeta_2$ is given by,
\begin{align}
\zeta_2 = \frac{1}{5} \left(\sqrt{5} \sqrt{20 Rh^{-2} Re^{-2} + 640 Rh^{-1} Re^{-3} + 5120 Re^{-4} + 3 } + 10 Rh^{-1} Re^{-1} +160 Re^{-2} \right).
\end{align}
Here $P_2 (t)$ is an oscillating complex quantity for the four-mode model or is a stationary real quantity for the three-mode model and in both cases $|P_2|$ is independent of time. This gives the following expressions for $\widetilde{\psi}_{\bf k}$ and $\widehat{\psi}_{\bf r}$, 
\begin{align}
\widetilde{\psi}_{\bf k} & = \frac{6}{16 Re^{-1} + Rh^{-1}} \zeta_2 | P_2 |^2, \label{eqn:sol_psik_app} \\
\widehat{\psi}_{\bf r} & = \frac{12}{4 Re^{-1} + Rh^{-1}} \zeta_2 | P_2 |^2. \label{eqn:sol_psir_app}
\end{align}
Now we need the expression of $\mathcal{V}^{-1}$ to solve Eq. \eqref{eqn:ampl_eqn_app}, its matrix form is denoted as
\begin{align}
\mathcal{V}^{-1} = [{\bf V}_1 \,\, {\bf V}_2]^{-1} = 
\left[
\begin{array}{cc}
\beta_1 & \beta_2 \\
- \beta_1 & \beta_3 \\
\end{array} 
\right] 
\label{eqn:exp_Vinv_app}
\end{align}
where $\beta_1, \beta_2, \beta_3$ are defined as
\begin{align}
\beta_1 & = \frac{i \sqrt{5}}{2 \sqrt{20 Rh^{-2} Re^{-2} + 640 Rh^{-1} Re^{-3} + 5120 Re^{-4} + 3}}, \\
\beta_2 & =  \frac{\sqrt{5} \sqrt{20 Rh^{-2} Re^{-2} + 640 Rh^{-1} Re^{-3} + 5120 Re^{-4} + 3}+10 Rh^{-1} Re^{-1} +160 Re^{-2}}{2 \sqrt{5} \sqrt{20 Rh^{-2} Re^{-2} + 640 Rh^{-1} Re^{-3} + 5120 Re^{-4} +3}}, \\
\beta_3 & = \frac{\sqrt{5} \sqrt{20 Rh^{-2} Re^{-2} + 640 Rh^{-1} Re^{-3} + 5120 Re^{-4} + 3}-10 Rh^{-1} Re^{-1} -160 Re^{-2}}{2 \sqrt{5} \sqrt{20 Rh^{-2} Re^{-2} + 640 Rh^{-1} Re^{-3} + 5120 Re^{-4} + 3}}. 
\end{align} 
Substituing the expressions for $\widetilde{\psi}_{\bf k}, \widehat{\psi}_{\bf r}$ from Eqs. \eqref{eqn:sol_psik_app}, \eqref{eqn:sol_psir_app} and the expression of $\mathcal{V}^{-1}$ from Eq. \eqref{eqn:exp_Vinv_app}, into the Eq. \eqref{eqn:ampl_eqn_app} gives the amplitude equations for the \sk{growing} eigenmode $P_2 (t)$.  
By setting $\widehat{\psi}_{\bf r} = 0$ in the nonlinear term $N$ of Eq. \eqref{eqn:ampl_eqn_app} we get the amplitude equation for the pitchfork bifurcation (see Eq. \eqref{eqn:threemode}). By considering all four modes, with $\widehat{\psi}_{\bf r}$ taken from Eq. \eqref{eqn:sol_psir_app}, the resulting amplitude equation is given by Eq. \eqref{eqn:fourmode} which leads to oscillations.

\bibliographystyle{unsrt}
\bibliography{references}

\end{document}